\documentclass[reprint,amsmath,amssymb,aps]{revtex4-1}
\usepackage{graphicx}
\usepackage{dcolumn}
\usepackage{bm}
\usepackage{braket}
\usepackage{subfigure}
\begin{document}
\preprint{APS/123-QED}
\title{Mechanical stretching of amyloid A$\beta_{11-42}$ fibrils using steered molecular dynamics}
\author{Ashkan Shekaari\textsuperscript{a,}}
\thanks{shekaari@email.kntu.ac.ir}
\author{Mahmoud Jafari\textsuperscript{a,}}
\thanks{Corresponding author}
\email{jafari@kntu.ac.ir}
\affiliation{\textsuperscript{a}Department of Physics, K. N. Toosi University of Technology, Tehran, 15875-4416, Iran}
\date{\today}
\begin{abstract}
Mechanical strength of amyloid beta fibrils has been known to be correlated with neuronal cell death. Here, we resorted to steered molecular dynamics (SMD) simulations to mechanically stretch a single S-shape amyloid beta A$\beta_{11-42}$ dodecamer fibril in vacuum. It was found that the weakest sites at which the fibril was ruptured due to mechanical extension were exclusively at the interfaces of alanine and glutamic acid distributed throughout the fibril. It was also revealed that the free energy required to unfold the fibril to form a long linear conformation is equivalent to $\sim 210$ eV, being several thousand times larger than thermal voltage at room temperature. As a consequence, within solution a larger free energy is needed for such a maximal stretching based on the fact that amyloid beta fibrils are structurally more stable in solution due to the interplay between their hydrophobic cores and solution's entropy.
\begin{description}
	\item[PACS numbers]
87.14.Ee, 87.15.He, 87.15.Aa, 87.19.Xx, 87.15.--v
	\item[Keywords]
Alzheimer, Amyloid beta fibrils, Mechanical stretching, Steered molecular dynamics, Free energy, Rupture sites
\end{description}
\end{abstract}
\maketitle
\section{\label{sec:1}Introduction}
Amyloid fibrils formed by misfolded protein aggregations have recently turned into a subject of extensive research due to their indubitable role in pathogenesis of a number of neurodegenerative diseases~\cite{1} particularly Alzheimer's disease (AD)~\cite{2} and Parkinson's disease (PD)~\cite{3}. These amyloid fibrils exhibit the structural feature in a way that they form one-dimensionally ordered structures~\cite{4}, being quite stable and not easily dissolved in physiological conditions~\cite{5}. Structural stability of amyloid fibrils has in fact been attributed to their $\beta$-sheet extended backbone conformation as a mechanically resistant protein component~\cite{6}. 

Intriguing mechanical properties of amyloid fibrils~\cite{7,8} are indeed comparable to those of other mechanically resistant, protein-based materials such as spider silk~\cite{9,10}. Atomic force microscopy (AFM) experiments~\cite{11,12,13,14} have measured elastic modulus of amyloid fibrils to be of order of 1 GPa. Computational investigations have also estimated this property with values within 1--10 GPa~\cite{15,16,17,18,19}. The fracture toughness of amyloid fibrils with a length-scale of 3 nm has also been calculated to be about 30 kcal$\big/$(mol.nm$^{3}$)~\cite{20}, which is comparable to that of spider silk protein crystal with a length-scale of 2 nm. 

Mechanical properties of amyloid fibrils are indeed of great importance, particularly by taking into account the observation that they directly affect fibrils' biological functions. As an illustration, mechanical disruption of the cell membrane caused by amyloid fibrils~\cite{21} is based on the fact that its elastic modulus ($\sim100$ kPa~\cite{22}) is three orders of magnitude smaller than those of amyloid fibrils~\cite{23}. Investigating mechanical properties of proteins and protein-based materials indeed provides a means to determine the pathways governing their unfolding/refolding processes when they are subject to mechanical traction. An optical tweezer-based force spectroscopy, for example, has been applied to investigate mechanical unfolding and refolding of a single prion protein (PrP) in order to provide a deeper understanding of its misfolding mechanism, which also directly contributes to the formation of amyloid oligomers as nucleation seeds~\cite{24}. 

Based on the importance of mechanical properties of proteins and bio-macro molecules particularly amyloid fibrils, the present work has therefore been devoted to modeling mechanical stretching of a single S-shape amyloid beta A$\beta_{11-42}$ dodecamer fibril~\cite{25} (Fig.~\ref{fig:1}) via resorting to steered molecular dynamics (SMD)~\cite{28} computer simulations in vacuum. As a result, a large set of SMD simulations have been prepared examining of which has consequently led us to a number of biologically important findings including the weakest sites at which the fibril is ruptured due to mechanical stretching, as well as the free energy of unfolding the fibril to a long linear conformation using mechanical forces. Sec.~\ref{sec:2} describes our applied computational setup in detail. 
\section{\label{sec:2}Computational details}
\begin{figure}[t]
	\includegraphics[scale=0.25]{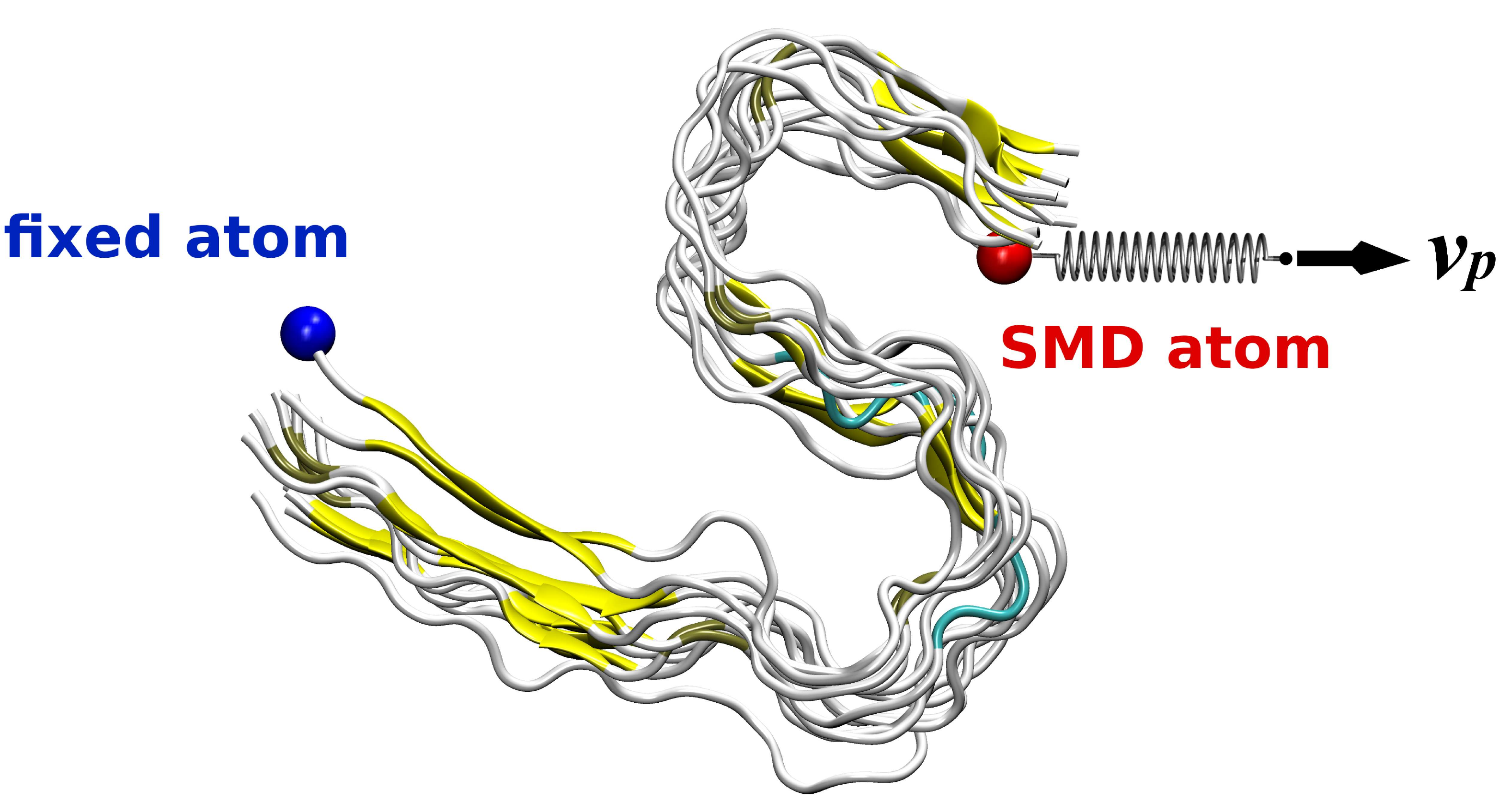}
	\caption{\label{fig:1} Tertiary structure of S-shape amyloid beta (A$\beta_{11-42}$) dodecamer fibril---rendered in VMD~\cite{38} using Tachyon parallel/multi-processor ray tracing system~\cite{39}. The fixed and SMD atoms at the two ends of the fibril are shown by the blue and red balls, respectively. The moving guiding potential applied in pulling experiments is also represented by a spring connected to C-terminus and is pulled with the constant velocity of $v_p$.}
\end{figure}

Constant-velocity SMD simulations have been applied to model mechanical stretching of a single S-shape amyloid beta dodecamer fibril (A$\beta_{11-42}$) in vacuum, as implemented in NAMD (version 2.14b2)~\cite{29} computational software. Initial atomic positions of the fibril have been taken from RCSB~\cite{30} Protein Data Bank (PDB) with entry code 2MXU~\cite{31}. Each simulation has been carried out in parallel, using Debian-style~\cite{33} Linux~\cite{34} operating system supported by MPICH (version 2-1.4)~\cite{35,36}. The July 2018 update of CHARMM36 force fields~\cite{32} has also been used. The fibril was solvated in a box of water under periodic boundary conditions with a unit-cell padding of 2.0 nm to decouple weakest periodic interactions. The minimization simulation was then carried out for 50000 conjugate-gradient steps (100 ps) followed by a 2-ns free-dynamics equilibration at 310 K and 1.01325 bar. Langevin forces with a damping constant of 2.50/ps along with Langevin piston pressure control has been applied for NPT equilibration to keep temperature and pressure of the system fixed. The subsequent SMD simulations have then been carried out in vacuum with both thermostat and barostat turned off to keep from disturbing atomic movements. The C$_\alpha$ atom of N-terminus tail located on the first residue (namely on glutamic acid with residue number 11) of the first chain A (all abbreviated as Glu$^{11}_{\mathrm{A}}$:C$_\alpha$) was kept fixed (Fig.~\ref{fig:1}), while Ala$^{42}_{\mathrm{L}}$:C$_\alpha$ (on last residue of last chain) has then been pulled along $+z$ perpendicular to fibril's axis, as illustrated in Fig.~\ref{fig:1}. A distribution of pulling velocities with values ranging from $10^{-2}$ to 10.0 \AA/ps were tested to reach the optimized value (namely $ v_{p}=0.1{\hspace{1mm}}{\mathrm{\AA}}/{\mathrm{ps}}=10{\hspace{1mm}}{\mathrm{nm}}/{\mathrm{ns}}$) for which the fibril was not easily ruptured. With this value, the pulling processes were also reversible in a way that the associated pulling work values were equivalent to free energy difference (${\mathrm\Delta} F$) between initial and final equilibrium states of the fibril. The force exerted on the SMD atom is described by
\begin{widetext} 
\begin{equation}
\label{eq:fr}
	{\bf{f}}({\bf{r}},t)=-{\bm{\nabla}} U({\bf{r}},t)=-{\bm{\nabla}}\left(\frac{1}{2}k_{s}\big[v_{p}t-({\bf{r}}-{\bf{r}}_{0}).{\bf{n}}\big]^{2}\right), 
\end{equation}
\end{widetext}
where $U$ is the harmonic potential (spring), $k_{s}$ is spring constant, $t$ is time, ${\bf{r}}$ and ${\bf{r}}_{0}$ are respectively the instantaneous and initial position vectors of SMD atom, and ${\bf{n}}=(0,0,1)$ is the normalized pulling vector along $+z$ (bold characters denote vectors). Having pulling velocity fixed, a set of twenty SMD simulations were first carried out with different values of $k_s$ ranging from 0.8 to 10.0 with an increment of 0.5 kcal$\big/$(mol.\AA$^{2}$) corresponding to thermal fluctuation of $\sqrt{kT/k_{s}}=1.11$ {\AA} for SMD atom at 310 K; $k$ being Boltzmann constant. Total accumulative time for this set of experiments is also 75.50 ns. As a result, maximum extension (53.387 nm) was found to take place for $k_{s}=2.0$ kcal$\big/$(mol.\AA$^{2}$), as seen in Fig.~\ref{fig:2}. 

The next set involving forty SMD experiments were then performed with both $v_{p}=0.1$ \AA/ps and $k_{s}=2.0$ kcal$\big/$(mol.\AA$^{2}$) to have a sufficient sampling of the maximum-extension experiment. The associated free energy $\Delta F$ required for realization of this process has accordingly been estimated using Jarzynski's equality~\cite{37}:
\begin{equation}
	\label{eq:1}
	e^{-\beta{\mathrm\Delta} F}=\Big\langle e^{-\beta W}\Big\rangle,
\end{equation}
where $\beta=1\big/kT$ is inverse temperature of the surrounding, and $W$ is the associated work being done during one realization of these (thermodynamically non-equilibrium) stretching processes from the initially unstretched (relaxed) state of the fibril to the point where it begins to rupture. These work values have been calculated via numerical integration of the corresponding time-dependent forces exerted on SMD atom using $W\simeq v_{p}{\mathrm\Delta} t\sum_{i} f(t_{i})$, where ${\mathrm\Delta} t=10$ fs. $x(t)=x_{eq}+v_{p}t$ is the distance between fixed and SMD atoms, and $x_{eq}=7.21$ nm at $t=0$ [the net elongation is therefore $x(t)-x_{eq}$].
The integration time-step is 1 fs. Hydrogen bonds have been identified within hydrogen donor-acceptor distance of 3.0 {\AA} and with cutoff angle 20$^{\circ}$. Solvent-accessible surface area (SASA, in units of nm$^2$)~\cite{40} has been calculated using the rolling-ball algorithm~\cite{41}, with a radius of 1.40 {\AA} for probe sphere. The VMD program has also been used to post-process and analyze the outputs.
\section{\label{sec:3}Results}
\begin{figure}[t]
	\includegraphics[scale=0.3]{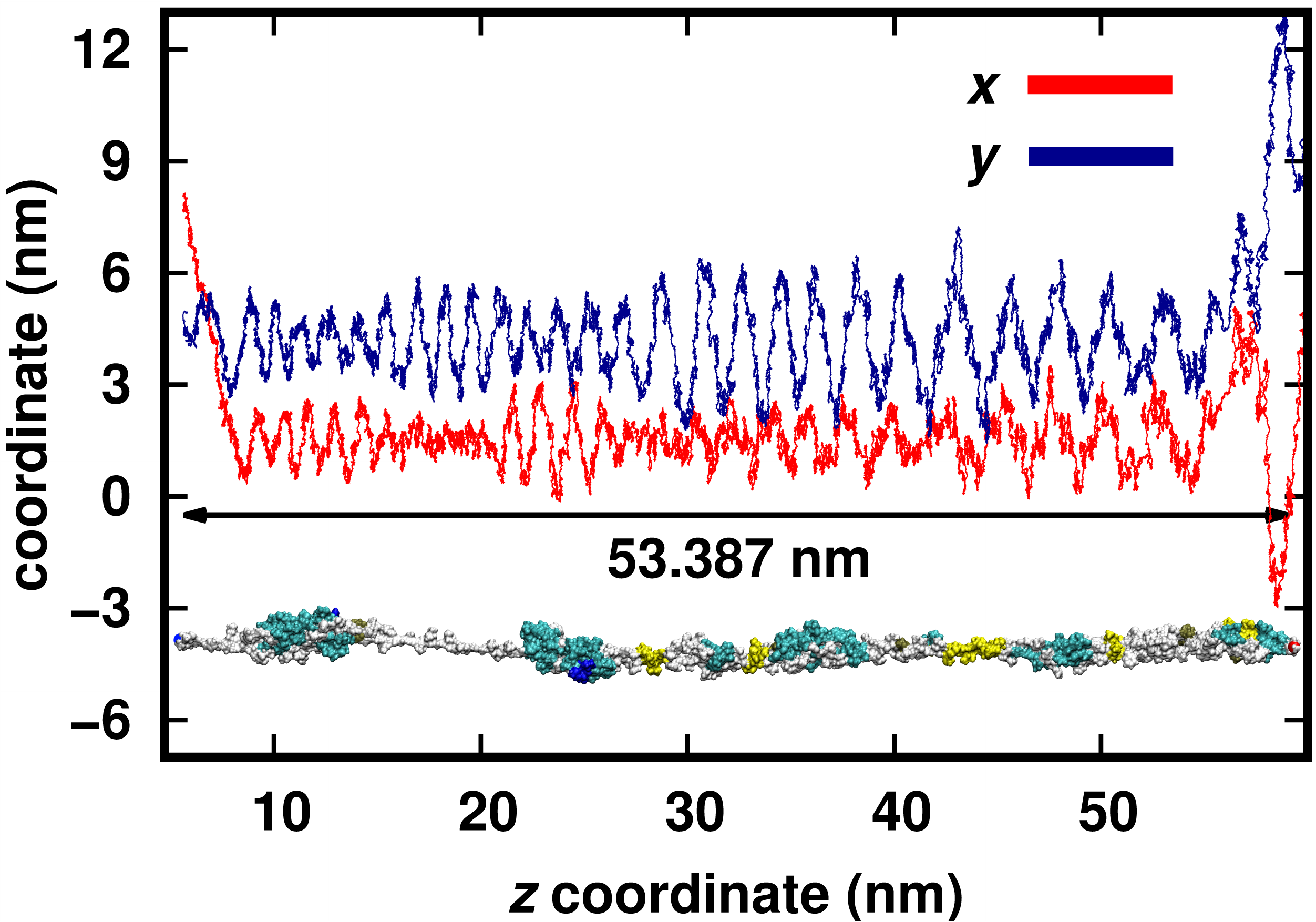}
	\caption{\label{fig:2} The trace of SMD atom in both $x-z$ (red) and $y-z$ (blue) Cartesian planes during the simulation with $k_{s}=2.0$ kcal$\big/$(mol.{\AA}$^{2}$) and $v_{p}=0.1$ \AA/ps---rendered in Gnuplot (version 5.2)~\cite{42}. The two zigzag patterns clearly arise from the S-shape structure of the twelve $\beta$-strands.}
\end{figure}

During equilibration, the fibril remained stable with a C$_{\alpha}$ root mean square deviation (RMSD) of less than 6.5 {\AA} and with all-atom RMSD of $<7$ {\AA} from the reference initial structure. The dummy atom attached to the SMD one via a spring with constant $k_s$ is pulled at $v_{p}=0.1$ \AA/ps along $+z$, making the latter atom to experience a force linearly dependent on $x(t)$. Here, we discuss the results obtained for a maximum-extension simulation [$k_{s}=2.0$ kcal$\big/$(mol.{\AA}$^{2}$)] with the extension length 53.387 nm, for which the trajectory of SMD atom in both $x-z$ and $y-z$ Cartesian planes during the simulation is shown in Fig.~\ref{fig:2}.
The fluctuating pattern is obviously a consequence of S-shape structure of the $\beta$-strands in $x-y$ plane, making SMD atom to accordingly move on a chiral path in three dimensions. The ever-decreasing part of the $z-x$ diagram from the beginning to where fluctuations begin corresponds to alignment of the end-to-end axis, connecting fixed and SMD atoms to each other, along the pulling direction ($+z$). During the simulation, SMD atom accordingly travels the net maximum distance of $\sim 46.177$ nm ($=53.387{\hspace{0.5mm}}{\mathrm{nm}}-7.21{\hspace{0.5mm}}{\mathrm{nm}}$).

\begin{figure*}
	\includegraphics[scale=1.15]{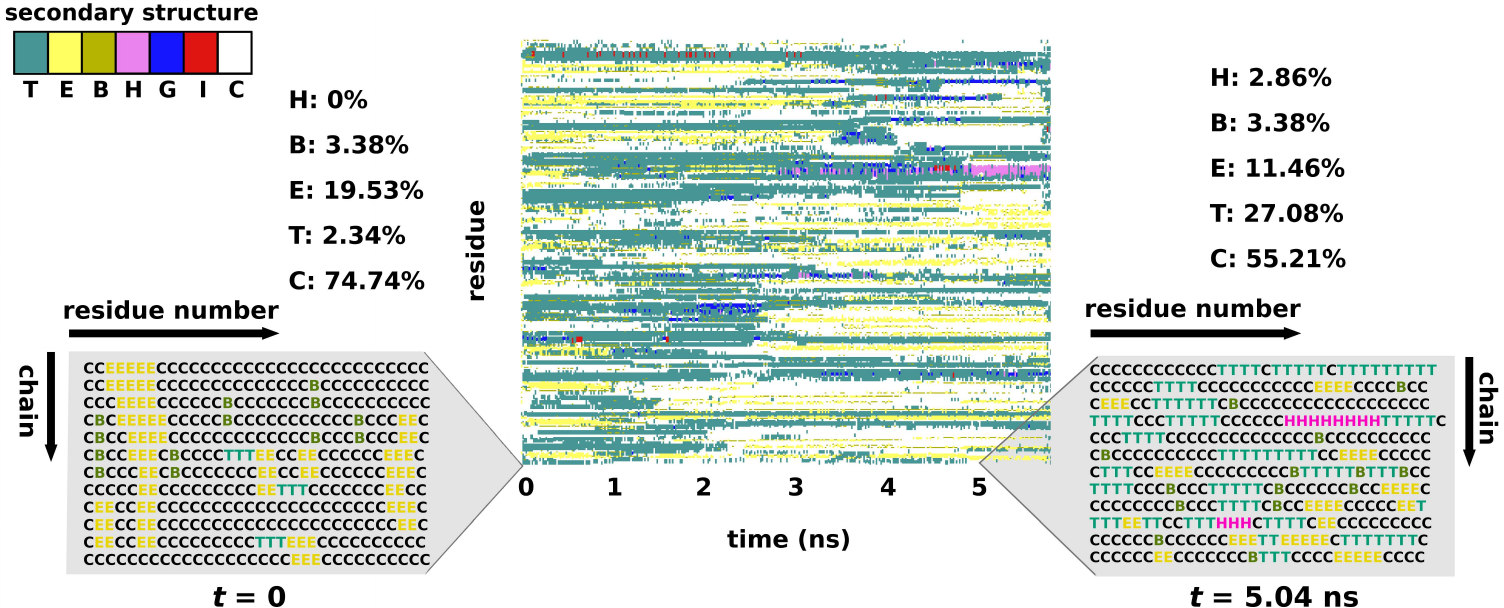}
	\caption{\label{fig:3}Secondary structure of the fibril during simulation--rendered by the Timeline plugin of VMD and edited by GIMP 2.8~\cite{43}. The two character-based representations correspond to secondary structure of the fibril equilibrated in solution ($t=0$) and at $t=5.04$ ns at which maximum elongation (53.387 nm) has been achieved. The (vertical) residue axis starts from Glu$_{\mathrm{A}}^{11}$ at the top and ends up to Ala$_{\mathrm{L}}^{42}$ at the bottom. T, E, B, H, G, I, and C, used by STRIDE algorithm~\cite{44} respectively stand for Turn, Extended, isolated Bridge, $\alpha-$Helix, 3$_{10}-$helix, $\pi-$helix, and Coil; their contributions to total secondary structure are also shown by the percent values. Rows and columns of the two character-based representations respectively show chains from A to L and residue numbers from 11 to 42.}
\end{figure*}

Fig.~\ref{fig:3} provides details about changes in secondary structure of the fibril.
As is seen, most of the diagram is filled by green and white areas, which respectively correspond to Turn and Coil. Examining the two character-based representations further reveals that during extension, Coil content decreases by 19.53\%, while that of Turn rises by 24.74\%. Extended content also reduces by 8.07\%, while that of $\alpha$-Helix increases from zero to 2.86\%. Comparing the representation associated to pre-equilibrated fibril in solution at $t=0$ and that of the initial crystal structure (the PDB file) also demonstrates that equilibration in water dramatically decreases Extended content from 62.5\% to 19.5\%.

\begin{figure}[h]
	\subfigure[]{\label{subfig:4(a)}
	\includegraphics[scale=0.7]{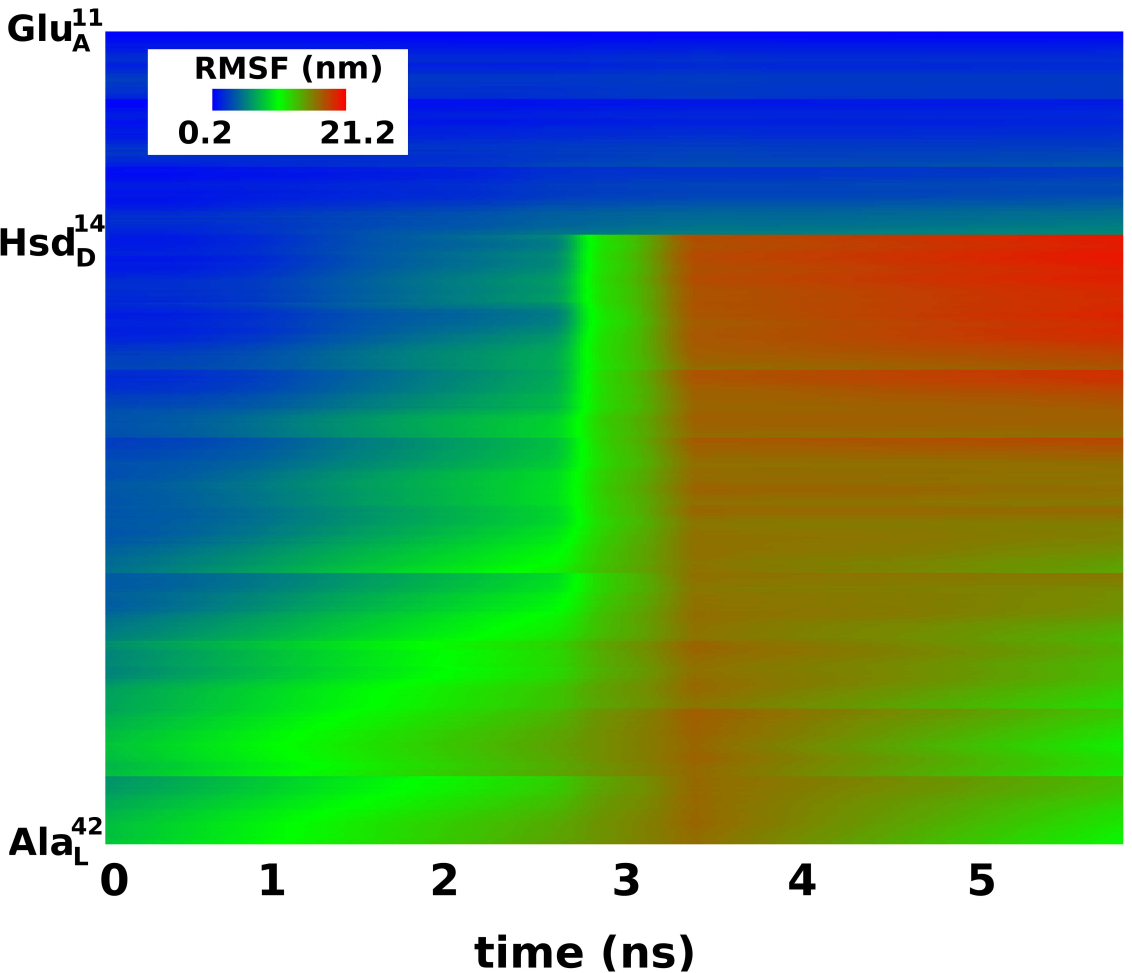}}
	\subfigure[]{\label{subfig:4(b)}
	\includegraphics[scale=0.7]{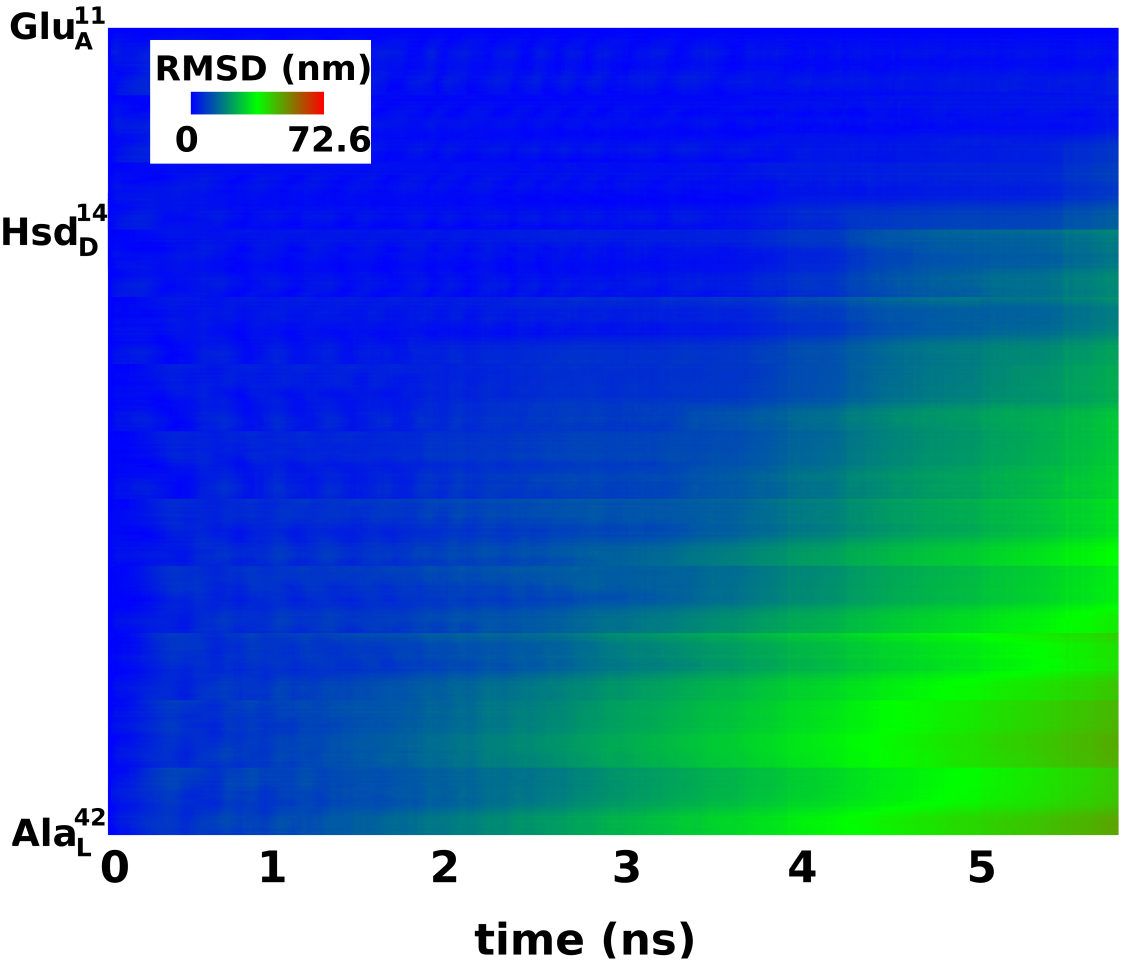}}
	\caption{\label{fig:4} Per-residue RMSF (a), and RMSD (b) of the fibril during the simulation. Dramatic changes begin from Hsd$^{14}_{\mathrm{D}}$ at about 2.5 ns.}
\end{figure}

Fig.~\ref{fig:4} shows per-residue root mean square fluctuation (RMSF) and RMSD of the fibril.
As is seen, the most varying (flexible) residues are those starting from Hsd$^{14}_{\mathrm{D}}$ (Hsd for histidine) to the last one (Ala$^{42}_{\mathrm{L}}$), as covered by the red region of Fig.~\ref{subfig:4(a)}. The blue region, from Glu$^{11}_{\mathrm{A}}$ to Hsd$^{14}_{\mathrm{D}}$, also indicates the least-flexible (rigid) residues during extension due to the fact that they are farthest amino acids from SMD atom. Fig.~\ref{subfig:4(b)} also shows these residues within the same range with nearly vanishing RMSD values as inferred from Fig.~\ref{subfig:4(a)}. RMSDs of the rest of the residues do not change over the first 2 ns; however from then on, they take an increasing trend in a way that the closer a residue is to SMD atom, the larger its RMSD.

\begin{figure*}
	\includegraphics[scale=0.7]{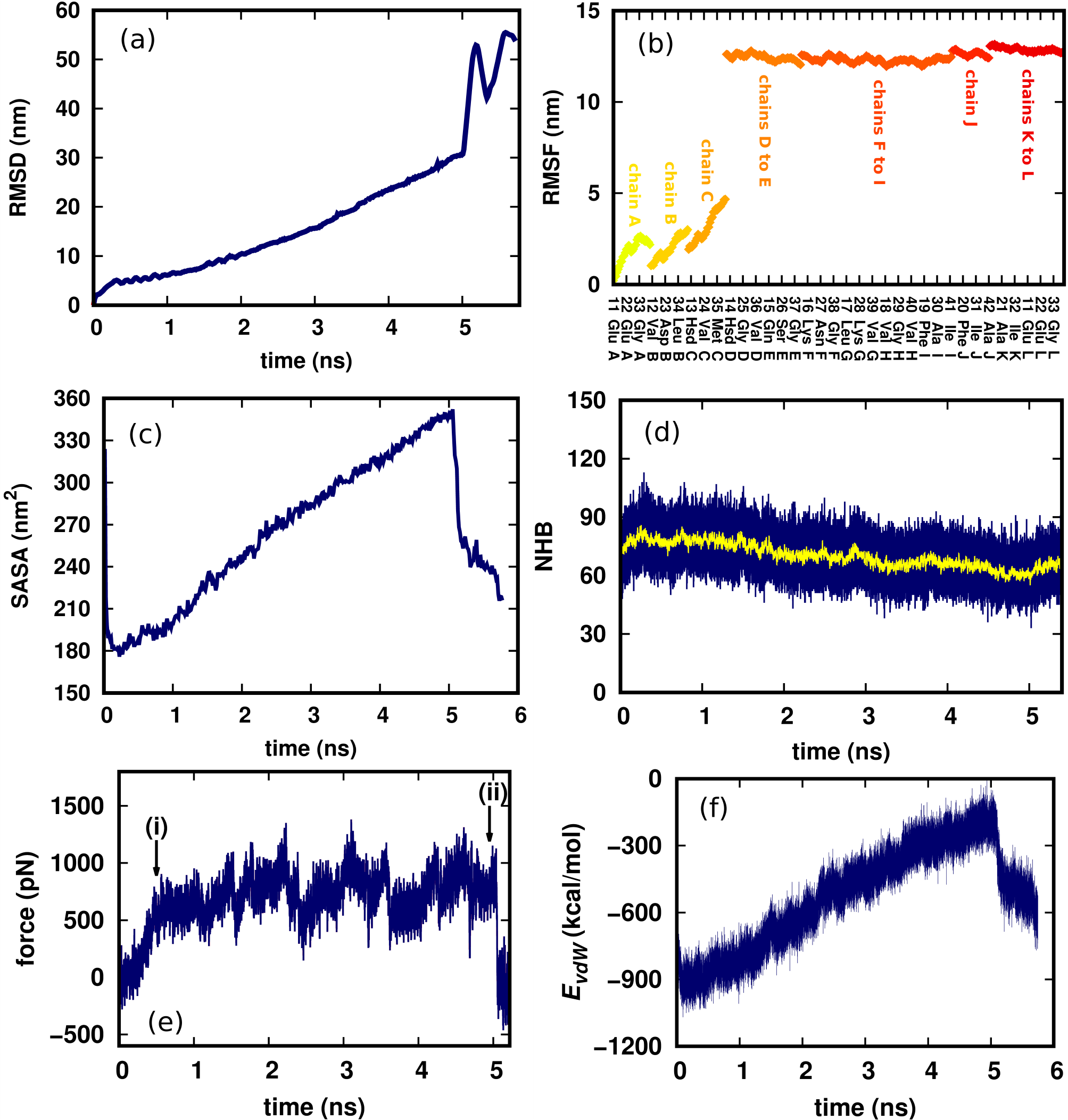}
	\caption{\label{fig:5} Time dependence of (a) total RMSD, (b) per-residue RMSF, (c) SASA, (d) number of hydrogen bonds (NHB), (e) force exerted on SMD atom according to Eq.~\ref{eq:fr}, and (f) $E_{vdW}$, over the entire trajectory. The instances labeled by arrows (i) and (ii) on the force-time profile respectively denote the points at which the fibril begins to stretch and when it finally approaches maximum elongation.}
\end{figure*}

Fig.~\ref{fig:5} also illustrates time dependence of a number of biologically important structural indicators of the fibril over the entire simulation time, as follows.
As is seen, the total RMSD curve takes an increasing trend up to 5.04 ns (corresponding to maximum extension) with a value of $\sim 31$ nm at this point, in agreement with RMSD of the green area in Fig.~\ref{subfig:4(b)}. From then on, however, an irregular zigzag pattern appears as a result of fibril's disruption. Fig.~\ref{fig:5}(b) exactly shows the mostly unchanged (rigid) as well as the mostly flexible residue ranges consistent with what were previously inferred from Fig.~\ref{subfig:4(a)}. The RMSF curve exhibits a drastic step-wise increase of $\sim 8$ nm at the place of Hsd$^{14}_{\mathrm{D}}$ (more exactly, at the interface of Ala$^{42}_{\mathrm{C}}$ and Glu$^{11}_{\mathrm{D}}$), which plainly reveals the weakest point (site) at which the fibril is ruptured. The parts with different colors further illustrate that the step-wise pattern, for other possible values of $k_s$ or $v_p$, would presumably take place at these joining points of fibril's chains. The chains K and L also exhibit largest RMSF values due to being nearest to SMD atom, as argued before.

Fig.~\ref{fig:5}(c) shows time dependence of SASA of the fibril during the simulation. It is seen that the time interval, from $t=0$ to global minimum of the diagram at $\sim 0.2$ ns corresponds to alignment of the end-to-end axis along $+z$. The why of this minimum in SASA is as such due to the fact that the fibril over this period is being relaxed and equilibrated in vacuum starting from the initial pre-equilibrated conformation in solution. Indeed, during equilibration in solution, water molecules have accordingly randomly been distributed around the fibril and more importantly within intrastrand regions (i.e., between the chains), which in turn, increase spacing between any two consecutive chains, leading to the value of $\sim 320$ nm$^2$ at $t=0$ as well. For $t>0$ in contrast, the solution has been removed and up to 0.2 ns during which stretching has not yet started, the fibril is equilibrated in vacuum and intrastrand spacing accordingly gets smaller. As a direct consequence, SASA takes a decreasing trend from $t=0$ on and is globally minimized at about 0.2 ns. The global maximum in the moving average of number of hydrogen bonds (NHB) [Fig.~\ref{fig:5}(d)] takes place exactly based on the same reasoning: removing water molecules from the intrastrand space makes chains get closer to each other and consequently increases NHB between them, maximizing NHB at minimum of SASA as well. In fact, variations of both SASA and NHB are intrinsically consistent. Extension up to 53.387 nm corresponds to increase/decrease in SASA/NHB with a nearly linear time dependence. Maximum of SASA also reveals the instance at which the fibril approaches its maximum length corresponding to minimum of NHB after which the fibril is to be ruptured.

Fig.~\ref{fig:5}(e) shows time dependence of the force exerted on SMD atom. Due to pulling, the fibril begins to rotate from $t=0$ on until the end-to-end axis is entirely aligned along the pulling direction ($+z$) at the instance denoted by arrow (i). From then on, extension fully starts during which acting bonds (e.g., hydrogen and van der Waals bonds, and salt bridges) in the fibril are constantly ruptured and reformed and so forth, leading to the saw-tooth pattern from (i) to (ii). The latter instance (ii) also corresponds to maximum elongation after which the fibril is ruptured and the force then immediately vanishes.

Examining time dependence of the van der Waals energy ($E_{vdW}$) of the fibril could also be interesting, as shown in Fig.~\ref{fig:5}(f). Similar to SASA, a three-stage pattern is seen: (1) decrease from $t=0$ to global minimum at about 0.2 ns; (2) increase from minimum to maximum of the diagram during which the fibril approaches its maximum extension; and (3) decrease from then on. The first decreasing trend obviously takes place according to the same reasoning provided before for explaining the time dependence of SASA [Fig.~\ref{fig:5}(c)]. At the minimum, the fibril is in fact in its most stable conformation due to being equilibrated in vacuum. 

During the second stage, however, the fibril becomes increasingly unstable with time; the maximum point evidently corresponds to maximum extension and therefore to least stability. After the fibril is ruptured at this point, each separated part of the fibril begins to refold again, and $E_{vdW}$ consequently seeks a second minimum. Taking into account the fact that $x$ (extension length) linearly depends on time [i.e., $x(t)=x_{eq}+v_{p}t$], the same brands of behavior could then be inferred for $E_{vdW}$ as a function of $x$. As a result, the minimum of $E_{vdW}(t)$ indicates an equilibrium state, which could then be fairly approximated by a simple harmonic oscillator, revealing the spring-like feature (reversibility) of the fibril under such mechanical extension experiments as well. In terms of statistical mechanics~\cite{45}, the more the fibril is stretched, the less the number of accessible conformations [i.e., microstates ($\mathrm\Omega$)]. The associated conformational entropy ($S$) of the fibril according to Boltzmann's formula ($S=k\ln\mathrm\Omega$) then becomes minimum/maximum at the maximum/minimum of $E_{vdW}(t)$.

\begin{figure}[h]
	\includegraphics[scale=0.3]{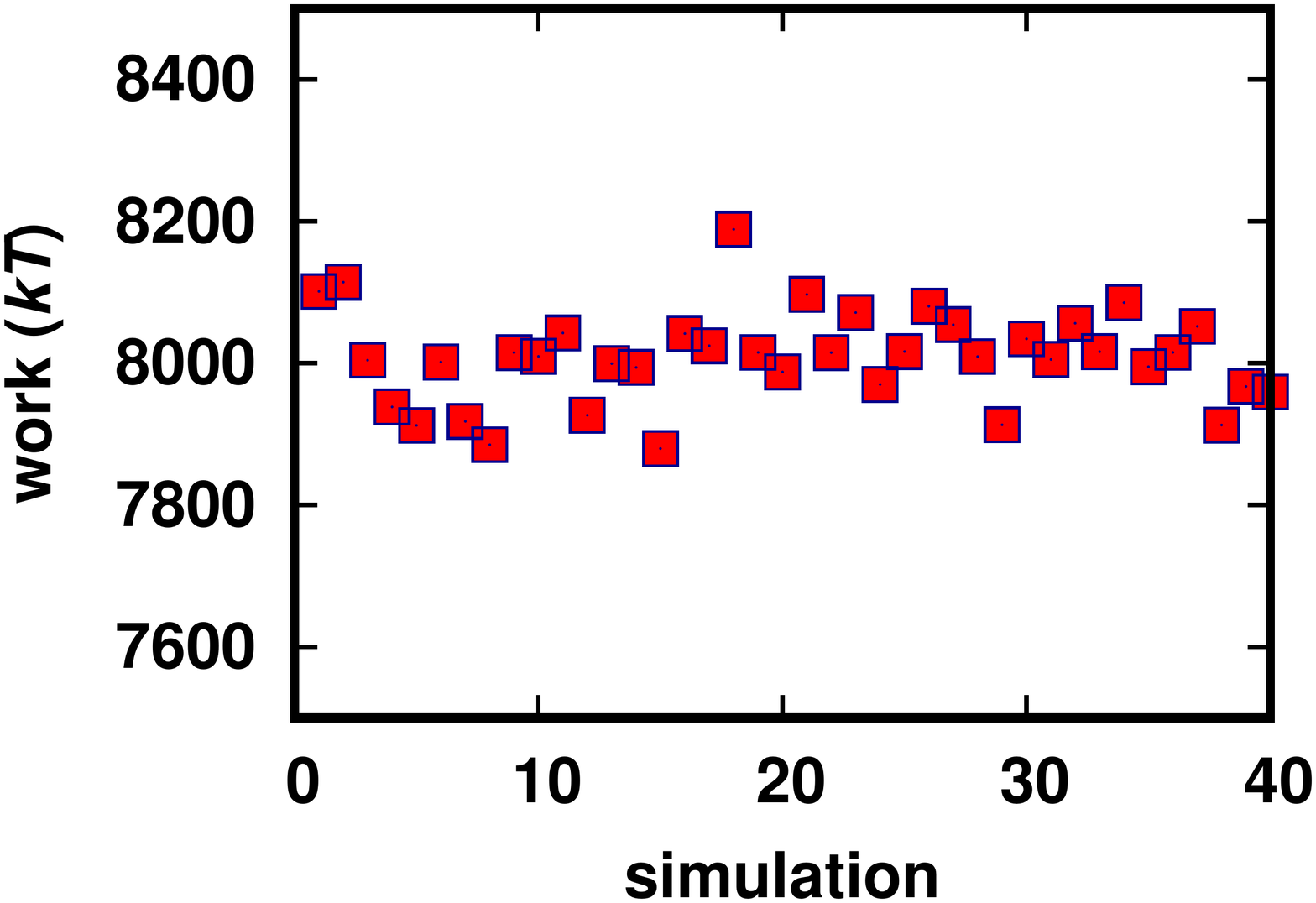}
	\caption{\label{fig:6} The work values calculated for our set of forty realizations of stretching simulation with $v_{p}=0.1$ \AA/ps and $k_{s}=2.0$ kcal$\big/$(mol.\AA$^{2}$), being distributed around the average 8008.119 in units of $kT$ at 310 K.}
\end{figure}

To estimate the free energy required to perform such maximal stretching experiments, we also have carried out a set of forty simulations all with $v_{p}=0.1$ \AA/ps and $k_{s}=2.0$ kcal$\big/$(mol.\AA$^{2}$), as their associated work ($W$) values are depicted in Fig.~\ref{fig:6} in units of $kT$ at 310 K.
Using these values, Eq.~\ref{eq:1} then results in
\begin{equation*}
\Delta F=-\ln\big\langle e^{-W}\big\rangle\simeq 7883\hspace{1mm}kT\simeq 210\hspace{1mm}{\mathrm{eV}},
\end{equation*}
which is equivalent to hydrolysis of $\sim 1735$ ATP (adenosine triphosphate) molecules; also is about 8108 times larger than thermal voltage at room temperature (0.0259 eV). Quite evidently, the probability that the fibril {\em{spontaneously}} unfolds to such maximally stretched states is zero, or technically, is equal to $e^{-\beta\Delta F}=e^{-7883}\sim 10^{-3424}$, meaning that the pathogenic conformation of amyloid fibrils is indeed irreversible and cannot be ever changed without the effect of an external agent.

Within solution, even a larger free energy is needed for such an extension based on the fact that enhancement in entropy of solvent (water) molecules expelled from the hydrophobic core between the two $\beta$-sheets of the fibril (hydrophobic effect) is key to fibril's stability~\cite{46}.

Finally to reveal the weakest sites at which the fibril is ruptured due to mechanical stretching, we carried out another set of simulations involving twenty SMD experiments in which $v_{p}=0.1$ \AA/ps but $k_s$ changes from 0.8 to 10 kcal$\big/$(mol.{\AA}$^{2}$), as described in Sec.~\ref{sec:2}. These sites has been tabulated in Table~\ref{tab:1}.
\begin{table}[!h]
	\caption{\label{tab:1}
		The weakest sites at which the fibril is ruptured due to mechanical traction. As is seen, these sites are exclusively at interfaces of alanine and glutamic acid (Ala--Glu) between different chains.}
	\begin{ruledtabular}
	\begin{tabular}{cl}
\raisebox{-0.7ex}{$k_s$ (kcal$\big/$mol.{\AA}$^{2}$)} &\raisebox{-0.7ex}{Rupture site}\\
\colrule
0.8 & \raisebox{-0.7ex}{Ala$^{42}_{\mathrm{J}}$--Glu$^{11}_{\mathrm{K}}$}\\
1.0 & \raisebox{-0.7ex}{Ala$^{42}_{\mathrm{F}}$--Glu$^{11}_{\mathrm{G}}$}\\
1.5 & \raisebox{-0.7ex}{Ala$^{42}_{\mathrm{I}}$--Glu$^{11}_{\mathrm{J}}$}\\
2.0 & \raisebox{-0.7ex}{Ala$^{42}_{\mathrm{C}}$--Glu$^{11}_{\mathrm{D}}$}\\
2.5 & \raisebox{-0.7ex}{Ala$^{42}_{\mathrm{E}}$--Glu$^{11}_{\mathrm{F}}$}\\
3.0 & \raisebox{-0.7ex}{Ala$^{42}_{\mathrm{I}}$--Glu$^{11}_{\mathrm{J}}$}\\
3.5 & \raisebox{-0.7ex}{Ala$^{42}_{\mathrm{I}}$--Glu$^{11}_{\mathrm{J}}$}\\
4.0 & \raisebox{-0.7ex}{Ala$^{42}_{\mathrm{A}}$--Glu$^{11}_{\mathrm{B}}$}\\
4.5 & \raisebox{-0.7ex}{Ala$^{42}_{\mathrm{J}}$--Glu$^{11}_{\mathrm{K}}$}\\
5.0 & \raisebox{-0.7ex}{Ala$^{42}_{\mathrm{B}}$--Glu$^{11}_{\mathrm{C}}$}\\
5.5 & \raisebox{-0.7ex}{Ala$^{42}_{\mathrm{E}}$--Glu$^{11}_{\mathrm{F}}$}\\
6.0 & \raisebox{-0.7ex}{Ala$^{42}_{\mathrm{I}}$--Glu$^{11}_{\mathrm{J}}$}\\
6.5 & \raisebox{-0.7ex}{Ala$^{42}_{\mathrm{B}}$--Glu$^{11}_{\mathrm{C}}$}\\
7.0 & \raisebox{-0.7ex}{Ala$^{42}_{\mathrm{I}}$--Glu$^{11}_{\mathrm{J}}$}\\
7.5 & \raisebox{-0.7ex}{Ala$^{42}_{\mathrm{A}}$--Glu$^{11}_{\mathrm{B}}$}\\
8.0 & \raisebox{-0.7ex}{Ala$^{42}_{\mathrm{G}}$--Glu$^{11}_{\mathrm{H}}$}\\
8.5 & \raisebox{-0.7ex}{Ala$^{42}_{\mathrm{E}}$--Glu$^{11}_{\mathrm{F}}$}\\
9.0 & \raisebox{-0.7ex}{Ala$^{42}_{\mathrm{I}}$--Glu$^{11}_{\mathrm{J}}$}\\
9.5 & \raisebox{-0.7ex}{Ala$^{42}_{\mathrm{G}}$--Glu$^{11}_{\mathrm{H}}$}\\
{\hspace{-1.7mm}}10.0& \raisebox{-0.7ex}{Ala$^{42}_{\mathrm{J}}$--Glu$^{11}_{\mathrm{K}}$}\\
		\end{tabular}
	\end{ruledtabular}
\end{table}
According to the table: (1) the weakest sites are exclusively at the interfaces of alanine (Ala) and glutamic acid (Glu) distributed throughout the fibril between different (consecutive) chains; (2) the weakest pair is the one between chains I and J due to being the most in number (contributing accordingly to 30\% of all Ala$^{42}$--Glu$^{11}$ pairs); and (3) other Ala--Glu contributions, in descending order, between different consecutive chains are 15\% for J--K and E--F, 10\% for A--B, B--C and G--H, and 5\% for C--D and F--G.
\section{\label{sec:4}Conclusions}
Steered molecular dynamics (SMD) simulations using the constant-velocity protocol have been applied to model mechanical stretching of S-shape amyloid beta A$\beta_{11-42}$ dodecamer fibrils in vacuum. A number of important structural indicators including RMSD, per-residue RMSF, number of hydrogen bonds, SASA, and force-time and energy-time profiles have accordingly been calculated and examined. A set of SMD simulations with varying spring constant $k_s$ has revealed that the weakest sites at which the fibril is ruptured due to mechanical stretching are exclusively at interfaces of alanine and glutamic acid (Ala--Glu) distributed over the entire fibrilbetween different consecutive chains. The weakest of them with the most contribution is also the one between chains I and J. Applying Jarzynski's equality to another larger set of SMD experiments with constant $k_s$ has also led to $\sim 210$ eV as an estimation of the free energy required to unfold the fibril to form a long linear conformation, being several thousand times larger than thermal voltage at room temperature. A larger free energy is accordingly required for such a stretching within solution (water) due to structural stabilizing effect of solution on the fibril arising from the interplay between fibril's hydrophobic core and solution's entropy.\\

\end{document}